\documentclass[aps,prb,preprint,groupedaddress]{revtex4}
\usepackage{graphicx}
\usepackage{amssymb}

\begin{document}


\title{Singlet ground state probed by crystal field inelastic neutron spectroscopy on the antiferroquadrupolar system
TmGa$_{3}$}

\author{M. G. Banks}
\email{m.banks@fkf.mpg.de}
\affiliation{Max-Planck-Institute f\"{u}r Festk\"{o}rperforschung,
Heisenbergstrasse 1, 70569 Stuttgart, Germany}

\author{R. K. Kremer}
\affiliation{Max-Planck-Institute f\"{u}r Festk\"{o}rperforschung,
Heisenbergstrasse 1, 70569 Stuttgart, Germany}

\author{D. Mannix}
\affiliation{ESRF, 6 rue Jules Horowitz, BP 200, F-38043, Grenoble
Cedex, France}

\author{G. Lapertot}
\affiliation{CEA-Grenoble, Department de Recherche Fondamentale sur
la Mati\`{e}re Condens\'{e}e 38054, Grenoble, France}

\author{A. P. Murani}
\affiliation{Institut Laue-Langevin, 6 rue Jules Horowitz, BP 156 -
F-38042 Grenoble Cedex, France }


\date{\today}

\begin{abstract}

TmGa$_{3}$ (AuCu$_3$ structure) undergoes two phase transitions, an
antiferroquadrupolar transition at $\sim$ 4.29 K and long-range
antiferromagnetic ordering at $\sim$ 4.26 K. Due to the close
vicinity of the two phase transitions, TmGa$_3$  offers an
interesting system to study the interplay of charge and magnetic
degrees of freedom.  In order to understand this interplay we have
performed inelastic neutron scattering experiments on TmGa$_{3}$ in
the paramagnetic regime ($T >$ 5 K) to redetermine the crystal
electric field level scheme. By fitting our spectra at various
temperatures we obtain a new crystal field level scheme with Lea,
Leask and Wolf parameters of $x_{\rm LLW}$ = -0.44(2) and $W$ =
-0.222(2) K.  The total crystal field splitting at 5K amounts to
$\sim$ 2.3 meV, about an order of magnitude less than found
previously, but in good agreement with the splitting extrapolated
from the related ErGa$_3$ system.

Our analysis yields a $\Gamma_{2}$ singlet as the crystal field
ground state followed closely by a (nonmagnetic) $\Gamma_{1}$
singlet at 0.009 meV. The next excited states are a
$\Gamma_{5}^{(2)}$ triplet at $\sim$0.5 meV, which is almost
degenerate to a $\Gamma_{4}$ doublet. This level scheme is adverse
to previous findings. Subsequent analysis of the magnetisation along
several crystallographic directions and the temperature dependant
susceptibility as well as of the magnetic contribution to the
specific heat are consistent with our new crystal field parameters.
Implications for the antiferroquadrupolar and the antiferromagnetic
transition are discussed.

\end{abstract}

\pacs{...}

\maketitle

\section{Introduction}

In 4\textit{f} electron systems the determination of the crystal
electric field (CEF) level scheme is an important first step to
understanding the rich variety of physical phenomena that rare earth
compounds exhibit. In 4\textit{f} systems the coupling between the
spin and orbital angular momentum is stronger than the crystal field
due to the fact that the 4\textit{f} orbital lies deep within the
ion core and the other occupied orbitals screen out the potential of
the surrounding ions. To a very good approximation the magnetic
properties of the free ion can therefore be characterized by the
total angular momentum \textit{J}. CEF interactions, in general,
lead to a splitting of the 2$J$+1 manifold into CEF states. Their
degeneracy and energy splitting is determined by the symmetry of the
electric charge distribution and the magnitude of the interaction.

Recently, there has been a renewed interest in systems which undergo
antiferroquadrupolar (AFQ) phase transitions originating from the
interaction of low lying quadrupole active CEF levels. CeB$_{6}$ was
the first compound to be unambiguously defined as having an AFQ
character by INS experiments.\cite{CeB6:Effantin} The phase diagram
of CeB$_{6}$ shows a huge enhancement of the AFQ phase with increase
of magnetic field aligned along [001]. More recently, a modulated
AFQ structure was identified in the intermetallic compound
PrPb$_{3}$ by INS experiments.\cite{PrPb:Onimaru} A non magnetic
$\Gamma_{1}$ ground state was found in the heavy fermion
superconductor PrOs$_{4}$Sb$_{12}$ which also undergoes a field
induced $O_{yz}$ type AFQ ordering.\cite{PrOs:Bauer,PrOs:Kaneko} The
heavy fermion system PrFe$_{4}$P$_{12}$ shows a large increase of
the lattice thermal conductivity  at the quadrupolar transition.
\cite{PrFe:Pourret} In UPd$_{3}$ a Q$_{zx}$ quadrupolar order
parameter was determined to describe the AFQ phase\cite{UPd:Walker}.






Quite a number of systems with AFQ transitions were found among Tm
based compounds. TmZn (Ref.\onlinecite{TmZn:luthi}), TmCu
(Ref.\onlinecite{TmCu:Morin}) and TmGa$_{3}$
(Ref.\onlinecite{Cz:tmspecheat}) exhibit an AFQ order in the
paramagnetic phase, followed closely by an antiferromagnetic (AFM)
transition. TmCd (Ref.\onlinecite{TmZn:luthi}), on the other hand,
shows only a transition of AFQ nature, with no experimentally
verified AFM transition down to 30 mK. TmTe, a magnetic
semiconductor which contains divalent Tm$^{2+}$ ions, has been
studied by INS, in which a field induced magnetic superstructure was
seen due to AFQ ordering at 1.8 K.\cite{TmTe:link} TmGa$_{3}$ was
intensively investigated due to its complex low temperature phase
diagram. The phase diagram of TmGa$_3$ was first investigated by
Czopnik \textit{et al.} by specific heat, thermal expansion and
susceptibility measurements.\cite{Cz:tmspecheat} They found two
close lying transitions at 4.26 K and 4.29 K. By a comparison with
TmZn and TmCd, the first transition at 4.29 K was attributed to  a
structural transition driven by quadrupolar pair interactions,
followed by an antiferromagnetic transition at 4.26 K. This
interesting interplay of charge and magnetic degrees of freedom
refocused our attention on TmGa$_{3}$. TmGa$_3$ crystallizes with
the AuCu$_{3}$ type cubic structure, space group
\textit{P}m$\bar{3}$m. The trivalent state, Tm$^{3+}$, with
electronic configuration 4\textit{f}$^{12}$ gives a total angular
momentum \textit{J} = 6. A first INS study carried out by Morin
\textit{et al.} \cite{Mo:tmnuetron} concluded the crystal field
parameters in the Lea, Leask and Wolf (LLW) scheme
(Ref.\onlinecite{Lw:leawolf}) to be \textit{x} = -0.32 and
\textit{W} = 1.03 K. These give a $\Gamma_{5}^{(1)}$ triplet  as the
CEF ground state separated by $\sim$2.5 meV from a $\Gamma_{3}$
state as the first excited state. A $\Gamma_{5}^{(1)}$ triplet CEF
ground state was reported to be consistent with the temperature
dependance of the magnetic susceptibility. A $^{169}$Tm
M\"{o}ssbauer spectroscopy study was used to determine the
quadrupolar term in the CEF Hamiltonian.\cite{PG:moss} By fixing
\textit{W} = 1.0 K, $x_{\rm LLW}$ = -0.38 and using an iterative
procedure the quadrupolar CEF term B$^{0}_{2}$ = -0.30 K was
obtained.

We decided to extend these preceding experiments and to search for
possible splittings of the $\Gamma_{5}^{(1)}$ ground state, as well
as follow the temperature dependance of the low crystal field levels
near the AFQ transition to see whether there are any signatures of
the AFQ transition in the CEF spectra. However, the results of our
inelastic crystal field spectroscopy experiments reveal a much
different CEF level scheme and overall splitting of the $J$=6
manifold in TmGa$_3$ that imply a reinterpretation of the low
temperature behavior of TmGa$_3$.

\section{Theoretical Background}

A prerequisite in the analysis of the magnetic properties of
TmGa$_3$ is the determination of the CEF ground state of the
Tm$^{3+}$ ions and the energy separation to the excited levels. For
a cubic system the Hamiltonian of the 2$J$+1 manifold of the total
angular momentum $J$ due to a crystal field interaction of cubic
symmetry is conveniently written in the operator equivalent form
\cite{Lw:leawolf}

\begin{equation}
\textsl{H}_{CEF} = B_{4}(\textsl{O}_{4}^{0} + 5 \textsl{O}_{4}^{4})
+ B_{6}(\textsl{O}_{6}^{0} - 21 \textsl{O}_{6}^{4}) \label{eq:cubic}
\end{equation}

where $\textsl{O}_{m}^{n}$ are the Stevens operators tabulated e.g.
in Ref. \onlinecite{aa:re}. The coefficients $B_{4}$ and $B_{6}$ are
parameters that measure the respective components in the multipole
expansion of the CEF potential. With the relations

\begin{equation}
B_4=A_4<r^4><J\|\beta\|J>
\end{equation}

and
\begin{equation}
B_6=A_6<r^6><J\|\gamma\|J> \label{A4A6},
\end{equation}

$B_4$ and $B_6$ can be further decomposed into a product of the
coefficients   $A_{4}$ and $A_{6}$, sometimes called geometrical
factors\cite{Lw:leawolf}, the mean fourth and sixth powers of the
radial part of the wave functions of the 4$f$ electrons, $<r^4>$ and
$<r^6>$ and the factors $<J\|\beta\,,\gamma\|J>$  which are listed
e.g. in Ref. [\onlinecite{aa:re}].

In order to conveniently tabulate the normalized eigenvectors and
eigenvalues for a given $J$ manifold,  Lea, Leask and Wolf suggested
a transformation of  $B_{4}$ and $B_{6}$ into the parameters $x_{\rm
LLW}$ and $W$ by using the following relations\cite{Lw:leawolf}

\begin{eqnarray}
B_{4}F(4) = W x_{\rm{LLW}} \\ B_{6}F(6) = W(1-|x_{\rm{LLW}}|)
\label{eq:b4b6}
\end{eqnarray}

where $F$(4) and $F$(6) are numbers given  in Ref.
\onlinecite{Lw:leawolf}.  Via this transformation the whole range
covered by $B_{4}$ and $B_{6}$ is mapped onto the the variable
$x_{\rm LLW}$ limited to the interval -1 $\leq x \leq$ +1 and the
energy scale factor $W$. With this transformation
eq.(\ref{eq:cubic}) now becomes

\begin{equation}
\frac{\textsl{H}_{CEF}}{W} = \left[ x_{\rm{LLW}}
\left(\frac{O_{4}}{F(4)} \right) +
(1-|x_{\rm{LLW}}|)\left(\frac{O_{6}}{F(6)} \right) \right]
\label{eq:cubicxw}
\end{equation}

with

\begin{eqnarray}
O_{4} = O^{0}_{4} + 5 O^{4}_{4} \\ O_{6} = O^{0}_{6} - 21 O^{4}_{6}
\label{eq:o4o6}
\end{eqnarray}

Empirically it was found that for a given system of compounds with
the same crystal structure the factors $A_4$ and $A_6$ vary only a
little across the rare earths series. \cite{dip:Walter,re:Walter}
Consequently, by using tabulated values for $<r^4>$ and $<r^6>$
(Refs.\onlinecite{re:Freeman,re:Brun}) and for $\beta_J$ =
$<J\|\beta\|J>$ and $\gamma_J$ = $<J\|\gamma|J>$ one can estimate
unknown parameters $B_4$ and $B_6$ from known CEF splitting found
for another rare earth ion in the series of isotypic compounds.

Morin \textit{et al.}, from the analysis of their INS spectra (see
above), found LLW parameters $x_{\rm LLW}$=-0.32 $\pm$ 0.02 and
$W$=1.03 $\pm$ 0.03 K  for TmGa$_3$, corresponding to $A_4<r^4>$ =
-34 $\pm$ 4 K and $A_6<r^6>$ = -17 $\pm$ 2 K with a $\Gamma_5^{(1)}$
triplet as the CEF ground state and an overall splitting  to the
uppermost $\Gamma_2$ singlet of 17.5 meV.\cite{Mo:tmnuetron}

A detailed analysis of the crystal field splitting of the Er$^{3+}$
manifold in the system ErGa$_3$ (isotypic to TmGa$_3$) was carried
out by Murasik \textit{et al.} by  inelastic neutron spectroscopy
and by the magnetic contributions to the heat capacity and the
magnetization\cite{Mu:erga3}. These experiments resulted in LLW
parameters for Er$^{3+}$ of ${x_{\rm LLW}}$ = +0.195 and ${W}$ =
+0.022 meV or equivalently, using eqs. (\ref{A4A6}) and
(\ref{eq:b4b6}), gives $A_4$(Er$^{3+}$) = +12.3 K and
$A_6$(Er$^{3+}$) = +3.85 K. Using the numerical values of
$<J\|\beta\|J>$ and $<J\|\gamma\|J>$ for Er$^{3+}$ and Tm$^{3+}$ and
also the values for $<r^4>$ and $<r^6>$ of Er$^{3+}$ and Tm$^{3+}$
respectively\cite{re:Freeman,re:Brun}, we find a large discrepancy
in the magnitude and sign, for the coefficients $A_4$ and $A_6$ for
Er$^{3+}$ in ErGa$_3$ to those of Tm$^{3+}$ in TmGa$_3$, as derived
from the result of  Morin \textit{et al.}.

TmGa$_3$ undergoes quadrupolar and antiferromagnetic ordering below
$\sim$4.3 K and $\sim$4.2 K, respectively (see above). Due to
magnetoelastic coupling, associated to these transitions is a
distortion of the lattice which leads to a symmetry lowering from
cubic symmetry.\cite{Cz:tmspecheat,PG:moss} This distortion induces
a further splitting of the CEF states that can be described by an
additional term added to the Hamiltonian in eq. (\ref{eq:cubicxw})

\begin{equation}
H_{QP} = B^{0}_{2}O^{0}_{2} + B^{2}_{2}O^{2}_{2}.
\label{eq:distortion}
\end{equation}

Quadrupolar-quadrupolar interaction between a pair of rare earth
ions labeled \textit{i} and \textit{j} is usually described by the
Hamiltonian \cite{aa:re,re:Birgeneau1966,re:Birgeneau1969}

\begin{equation}
{\textsl{H}_{QQ}} = A \left[ 4 O^0_{2i}\,O^0_{2i}-16
\left(O^{+1}_{2i}\,O^{-1}_{2i}+O^{-1}_{2i}\,O^{+1}_{2i} \right)+
\left(O^{+2}_{2i}\,O^{-2}_{2i}+O^{-2}_{2i}\,O^{+2}_{2i} \right)
\right ] \label{eq:quadquad}
\end{equation}

where the operators $O^l_k$ are defined in the usual way by
\begin{equation}
O^0_2=3J^2_z - J \left(J+1 \right),
 \label{eq:O20}
\end{equation}

\begin{equation}
O^{\pm 1}_2=\frac{1}{2}\left( J_zJ_{\pm } + J_{\pm }J_z \right),
 \label{eq:O21}
\end{equation}
and
\begin{equation}
O^{\pm 2}_2=J^2_{\pm }.
 \label{eq:O22}
\end{equation}

The coefficient \textit{A} is given by

\begin{equation}
A=\frac{3e^2<r^2_i><r^2_j><J_i||\alpha||J_i><J_j||\alpha||J_j>}{8\epsilon_{ij}
R^5}
 \label{eq:A}
\end{equation}

with $<r^2_i>$ being the mean square radius of the 4$f$ ions on each
site and $<J||\alpha||J>$ the coefficient listed e.g. in Ref.
\onlinecite{aa:re}. $R$ is the distance between the ions and
$\epsilon_{ij}$ is an effective dielectric constant for the pair of
ions. \cite{re:Birgeneau1966} Knowing $<r^2>$ and $\epsilon_{ij}$,
the coefficient $A$ can, in principle, be  calculated. However,
$\epsilon_{ij}$ is very sensitive to shielding effects due to the
intervening ions and the conduction electrons in the case of metals.
In general, $A$ must therefore be treated as an unknown parameter to
be obtained from experiment.

First order quadrupolar effects for Tm$^{3+}$ may occur for the
non-Kramers doublet $\Gamma_3$ with the two states having a
quadrupole moment of opposite sign but the same magnitude and for
the triplet states $\Gamma_4$ and $\Gamma_5$. The two singlets
$\Gamma_1$ and $\Gamma_2$ show only second and higher order
quadrupolar effects.

\section{Experimental}

All samples were grown  from a Ga flux in quartz ampoules with a
procedure  described elsewhere\cite{pc:tmgagrowth} and characterized
by X-ray diffraction to  ensure phase purity. They were  found to
have the correct AuCu$_{3}$ structure with no impurity reflections
above the noise level.  The composition of  single crystals was
checked by electron microprobe analysis and found to have the
correct atomic ratios.

Inelastic neutron CEF spectroscopy was performed on the neutron
time-of-flight spectrometers IN4, IN6 and the triple axis
spectrometer IN12 at the Institut Laue-Langevin (Grenoble). IN4
operates in the thermal neutron energy range 10-100 meV. A suitable
energy is selected from the thermal spectrum with a crystal
monochromator. The cold neutron time-of-flight spectrometer IN6
provides quasielastic and inelastic spectra for incident wavelengths
in the range of 4 to 6 {\AA}. A graphite monochromator delivers the
four wavelengths 4.1, 4.6, 5.1, and 5.9 {\AA}. The second order
reflection from the graphite monochromator is removed by a
beryllium-filter cooled to liquid nitrogen temperature. The elastic
energy resolution at 4.1 {\AA} is 170 $\mu$eV and at 5.9 {\AA}, 50
$\mu$eV. Inelastic neutron scattering measurements on a
polycrystalöline sample of Tm$_{0.1}$Lu$_{0.9}$Ga$_{3}$ were carried
out on the cold neutron triple-axis spectrometer IN12. The analysis
of the INS was carried out using the program $Mcphase$
\cite{MR:mcphase}  to calculate at a given temperature the
intensities and energies of the allowed CEF transitions for a set of
LLW parameters $x_{\rm LLW}$ and $W$.

In addition, we have performed specific heat measurements on single
crystals and polycrystalline samples of TmGa$_{3}$ and
polycrystalline samples of  Lu$_{1-x}$Tm$_{x}$Ga$_{3}$ ($x$=0.05,
0.01) and on the isostructural nonmagnetic compound LuGa$_{3}$ in a
PPMS (Quantum Design) relaxation-type calorimeter. The sample pieces
were fixed with a minute amount of Apiezon N grease on a sapphire
platform which also carried the heater and a calibrated temperature
sensor. The addenda heat capacities of the calorimeter and the
grease were determined in separate runs and subtracted. Magnetic
susceptibility measurements were performed using a MPMS magnetometer
(Quantum Design) in the temperature 1.9K $< T <$ 350K and 0$<$T$<B<$
7T.

Electron paramagnetic resonance (EPR) spectroscopy on Er$^{3+}$ in
polycrystalline samples with composition  Lu$_{1-x}$Er$_{x}$Ga$_{3}$
($x$=0.005, 0.01, 0.03) was performed using a Bruker  X-band EPR
spectrometer ($\nu \sim$9.3 GHz). The derivative of the resonance
absorption \textit{d}\textit{P}$_{abs}$/\textit{d}\textit{H} is
obtained using field modulation ($\nu$=100 kHz) with standard
lock-in detection technique. Coarse powdered samples were filled in
quartz ampoules and fixed in paraffin for measurements below room
temperature, which was provided by a continuous-flow He cryostat
(Oxford instruments).

\section{Results and Discussion}

\subsection{EPR on Lu$_{1-x}$Er$_{x}$Ga$_{3}$}

In order to test whether the finding of a $\Gamma_7$ doublet CEF
ground state for the $J$ = 15/2 manifold of Er$^{3+}$ in ErGa$_3$
can be confirmed, we carried out an EPR study on
Lu$_{1-x}$Er$_x$Ga$_3$ ($x$=0.005, 0.01, 0.03) in the temperature
range 4 K $< T $ $\\lesssim$25 K. A $\Gamma_7$ Kramers doublet is
expected to exhibit an isotropic EPR line at a resonance field
corresponding to a $g$ factor of $g(\Gamma_7) \approx$ 6.8. The
resonance field of the $\Gamma_7$ doublet can be well distinguished
from that of the $\Gamma_6$ doublet which has a $g$ factor of
$g(\Gamma_6) \approx$ 6.0. EPR resonances of the three
$\Gamma_8^{(1-3)}$ quadruplets resulting from the CEF splitting of
the $J$ = 15/2 manifold are angular dependent and in a
polycrystalline sample will generally result in very broad smeared
resonance lines. Excited  $\Gamma_8^{}$ CEF states at an energy
$\Delta \sim k_BT$  open additional relaxation channels via Hirst
and Orbach processes which give an exponential increase of the
linewidth, $\propto exp(-\Delta/k_BT)$) in addition to the normal
Korringa relaxation, $\propto T$.\cite{epr:hirst,epr:orbach1961}

Murasik \textit{et al.}'s CEF scheme provides a $\Gamma_8^{(1)}$
quadruplet as the first excited state at an energy of $\sim$2.7
meV.\cite{Mu:erga3} Measurements of the temperature dependence of
the EPR linewidth above liquid He temperature should be able to
reveal relaxation via this state and provide an additional support
for the validity of the proposed CEF level scheme of ErGa$_3$.

\begin{figure}
\includegraphics[bb = 17 14 270 235]{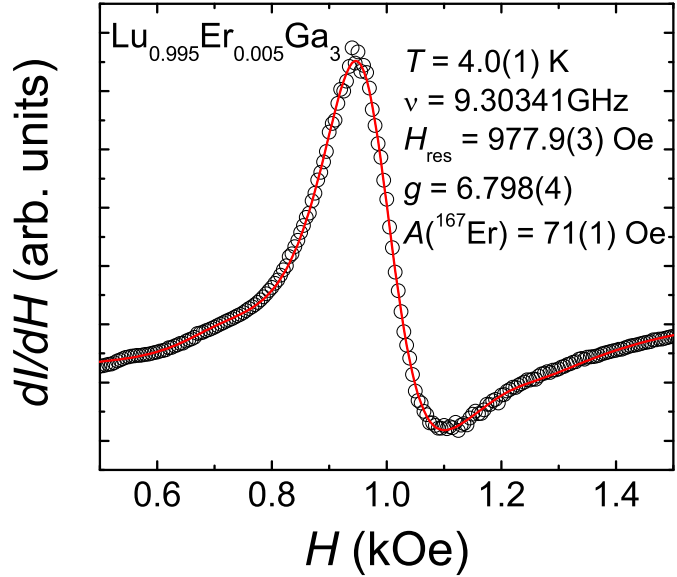}
\caption{(color online) $\circ$ EPR resonance line of Er$^{3+}$ in
the sample Lu$_{0.995}$Er$_{0.005}$Ga$_3$ at 4 K at a microwave
frequency of 9.3034 GHz. The (red) solid line is a fit with a
superposition of a central line and eight satellites (see text).
Their resonance position has been calculated with the Breit-Rabi
formula (Ref. \onlinecite{epr:pake,epr:Low}) with a hyperfine
constant $A(^{167}$Er)=71(1) Oe.} \label{fig:EPR}
\end{figure}

As a characteristic spectrum, Fig. \ref{fig:EPR} displays the EPR
resonance of Lu$_{0.995}$Er$_{0.005}$Ga$_3$ at a frequency of
$\sim$9.3 GHz. At low temperatures we observe an asymmetric
resonance line at resonance fields of $\sim$ 978(1) Oe with typical
linewidths of $\sim$100 Oe. With increasing temperature the
resonance  rapidly broadens and it is not detectable any more above
$\sim$25 K. As is typical for EPR  of localized moments in metals
and intermetallic compounds, the resonance is asymmetric due to the
mixtures of an absorption and dispersion signal due to the skin
effect. Characteristic shoulders, especially resolved for small Er
concentrations in the low field wing of the central line are due to
hyperfine satellites from the isotope $^{167}$Er ($I$ = 7/2). Their
resonance fields $H_{res}(m=-7/2, ...,7/2)$ shift with respect to
the resonance field of the central line, $H_0$, observed for all
other isotopes with $I$=0 according to:\cite{epr:Low}

\begin{equation}
H_{res}(m)=H_0 - ^{167}Am[1 + (\frac{^{167}A}{2H_0})^2] -
\frac{^{167}A^2}{2H_0}[I(I+1) - m^2]
 \label{eq:BreitRabi}
\end{equation}

where $^{167}A$ is the hyperfine constant for $^{167}$Er.

 A superposition of the
central line and the hyperfine satellites, all with the same
linewidth, and an absorption/dispersion ratio (characterizing the
linewidth asymmetry) of  typically $\sim$1 was found to fit the
spectra very well (cf. Fig. \ref{fig:EPR}).

The resonance fields of the central lines is independent of the
temperature within error bars and  corresponds to a $g$ factor of

\begin{equation}
g = 6.798(4).
 \label{eq:g}
\end{equation}

The $g$ factor and hyperfine constant $^{167}A$ are somewhat larger
than the respective quantities observed for the Er EPR of a
$\Gamma_7$ CEF doublet in insulators (6.75 - 6.76 and 73 - 74
G).\cite{epr:Tao} These shifts can be understood as due to the
exchange interaction of the localized Er moment with the conduction
electrons. \cite{epr:Tao,epr:barnes}

\begin{figure}
\includegraphics[bb = 17 14 303 227]{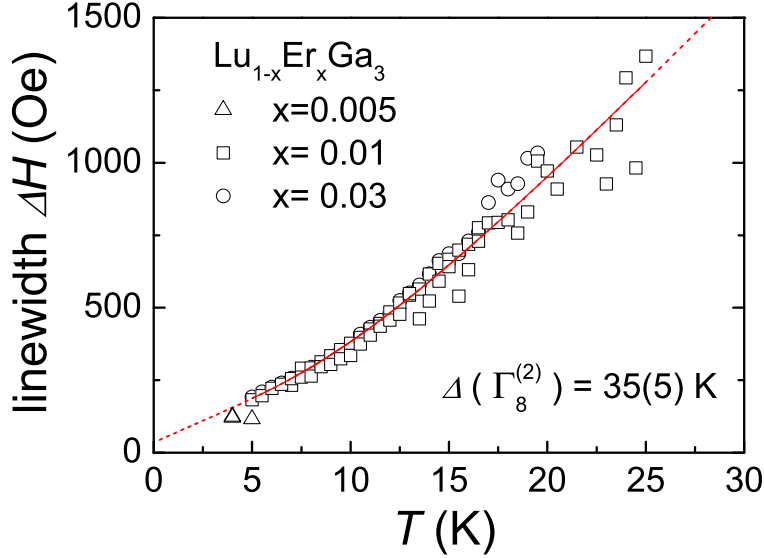}
\caption{(Color online) EPR linewidth $\Delta H$ as a function of
temperature for Lu$_{1-x}$Er$_x$Ga$_3$ with $x$=0.005, 0.01 and
0.03. The supralinear increase towards higher temperatures can be
fitted (dashed line) assuming an exponential increase according to
eq. (\ref{eq:eprline}) with an energy gap from the ground state to
the first excited state of $\Delta$ = 35 K or 3.1 meV.}
\label{fig:EPRlinewidth}
\end{figure}

The temperature dependence of the linewidth obtained from the fits
are displayed in Fig. \ref{fig:EPRlinewidth}. At low temperatures we
observe a linear increase of the linewidth with temperature
(Korringa relaxation). Towards higher temperatures the linewidth
grows faster than linearly and above $\sim$ 20 K the linewidth is of
the order of the resonance fields and  the EPR is not detectable any
more.

The non-linear temperature dependence of the EPR linewidth of
non-$S$ ground state rare earth  moments in metals can be
conveniently described by the function

\begin{equation}
\Delta H = \Delta H_{res} + b T + \frac{c}{exp(\Delta/k_{B}T)-1}
 \label{eq:eprline}
\end{equation}

where $\Delta H_{res}$ represents the residual linewidth, $b\,T$ the
Korringa relaxation of the localized moments to the conduction
electrons and the supralinear third term summarizes the contribution
from relaxation processes via excited CEF states located at an
energy $\Delta$ above the CEF ground state (Orbach and Hirst
processes).\cite{epr:orbach,epr:hirst} Fitting our data with eq.
(\ref{eq:eprline}) we arrive at a Korringa term

\begin{equation}
b =  30(2) {\rm {Oe/K}}
 \label{eq:b}
\end{equation}

The exponential broadening of the linewidth at higher temperatures
can be ascribed to relaxation via an excited CEF state at an energy

\begin{equation}
\Delta = 3.1(5) {\rm {meV}}
 \label{eq:delta}
\end{equation}

The g factor  of the Er$^{3+}$ EPR in Lu$_{1-x}$Er$_x$Ga$_3$ clearly
confirms the $\Gamma_7$ doublet as the CEF groundstate  and the
analysis of the temperature dependence of the linewidth sets the
first excited state to $\sim$3.1 meV, in very good agreement with
the energy of the first excited quadruplet concluded by Murasik
\textit{et al.}. These findings strongly support the analysis by
Murasik \textit{et al.} for the CEF level scheme of Er$^{3+}$ in
ErGa$_3$. They also show that by dilution into the isotypic
diamagnetic matrix, LuGa$_3$, the single-ion CEF parameters remain
close to those found in the concentrated compound  in the
paramagnetic state.

EPR spectra taken at low temperatures on polycrystalline samples of
Lu$_{0.95}$Tm$_{0.05}$Ga$_3$ showed no indication of a resonance
from Tm$^{3+}$.

\subsection{Inelastic Neutron Spectroscopy}

In their early INS investigation Morin \textit{et al.} observed
broad inelastic modes centered at $\sim$2.6 meV,  $\sim$11meV and
$\sim$17 meV from which they concluded a total splitting of the CEF
levels of $\sim$200K.\cite{Mo:tmnuetron} Guided by these results we
studied polycrystalline samples of TmGa$_3$,
Lu$_{0.95}$Tm$_{0.05}$Ga$_3$ and ErGa$_3$ on ILL's time-of-flight
spectrometer IN4 with neutrons of incident energies corresponding to
wavelengths of 3.3 \rm{\AA}, 2.2 \rm{\AA} and 1.1 \rm{\AA} at
temperatures of 20 K and below. Our data for TmGa$_3$ show modes at
the same energies as found by Morin \textit{et al.} but the mode at
$\sim$11meV is somewhat better resolved and it becomes  obvious that
this mode consists of two modes at $\sim$8.5meV and $\sim$10.7meV.
At the same energies, modes of comparable (normalized) intensity are
also observed in the IN4 spectra of ErGa$_3$ and of
Lu$_{0.95}$Tm$_{0.05}$Ga$_3$. In ErGa$_3$ Murasik \textit{et al.}
have seen inelastic scattering intensity emerging from CEF
transitions essentially up to $\sim$5meV.\cite{Mu:erga3} This energy
limitation for the CEF transitions in ErGa$_3$  and the fact that in
the diamagnetically diluted sample Lu$_{0.95}$Tm$_{0.05}$Ga$_3$
similar modes are observed, leads us to the conclusion that all
modes appearing in TmGa$_3$, ErGa$_3$ and
Lu$_{0.95}$Tm$_{0.05}$Ga$_3$ above  $\sim$6meV cannot be attributed
to CEF transitions. The subsequent inspection of the
\textit{\textbf{q}}-dependence of these modes clearly identified
them as phonon modes. Extending the energy range further, two more
phonon modes at $\sim$21.5meV and $\sim$35meV were found.

The detailed investigation of the temperature dependence of the low
energy regime, $E_i<$7.5meV, showed that the mode centered at
$\sim$2.6meV consists at least of two submodes and revealed also a
mode at very low energy $\lesssim$0.4meV appearing as a shoulder on
the energy-loss side of the elastic peak. Comparison with the
energy-gain part of the spectra indicated this mode to originate
from a CEF excitation from states close to the ground state.
Summarizing these results, we concluded that, in contrast to the
preceding analysis by Morin \textit{et al.} the total CEF splitting
of the $J$=6 manifold of Tm$^{3+}$ in TmGa$_3$ does not exceed an
energy range of $\sim$3.5meV. This presumption was in agreement with
the energy range estimated from the $A_4$ and $A_6$ parameters
obtained from the analysis of the ErGa$_3$ INS data (see above).
Using the IN4 data a least square fitting procedure was carried out
on the 5 and 10 K spectra (above the phase transitions). Although
the 0.3meV excitation is not clearly resolved, the relative
intensities of the peaks at 0.3meV and 2.5meV enabled us to obtain a
first estimate of the LLW parameters which were in accord with this
conclusion.

\begin{figure}
\includegraphics{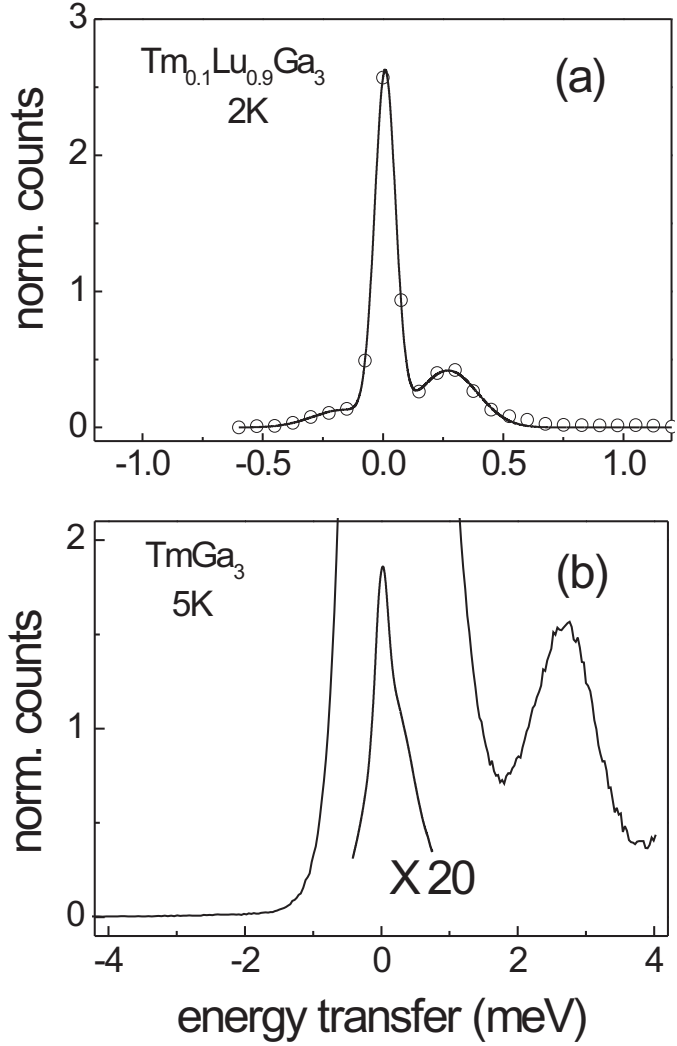}
\caption{INS spectra ($\circ$) of (a) Lu$_{0.9}$Tm$_{0.1}$Ga$_3$
collected on the triple-axis spectrometer IN12 at 2K (the solid line
is a guide to the eye) and of (b) TmGa$_3$ collected on the TOF
spectrometer IN6 at 5K.} \label{fig:INS12lin}
\end{figure}

In order to better resolve the energy regime below $\sim$4 meV, we
performed measurements using lower incident energies on  the cold
source TOF spectrometer IN6. To further improve the resolution in
the regime below $\sim$1 meV, spectra of the diluted system
Lu$_{0.9}$Tm$_{0.1}$Ga$_3$ were taken in the cold neutron regime on
ILL's triple-axis spectrometer IN12. Fig.\ref{fig:INS12lin} displays
a typical INS spectrum of TmGa$_3$ measured on IN6 at 5K in the
paramagnetic regime and of Lu$_{0.9}$Tm$_{0.1}$Ga$_3$ obtained with
the triple-axis spectrometer IN12 at 2K. The spectrum of TmGa$_3$
clearly shows a broad structured mode centered at 2.6meV with a FWHM
of $\sim$0.8meV, larger than the experimental resolution. This
broadening indicates that this mode consists of several overlapping
lines, as will be discussed in detail below. A strong mode
overlapping with the elastic line on the energy-loss side is now
well resolved. It becomes more clearly visible as a mode separated
from the elastic line in the IN12 spectra of
Lu$_{0.9}$Tm$_{0.1}$Ga$_3$ at 2K.

To fit the INS spectra we used the program $Mcphase$
\cite{MR:mcphase} which provides  the INS total scattering cross
section (in barn/ion) for a transition from level $n$ to $m$ at a
set temperature with the partition function $Z$ and with LLW
parameters $x_{\rm LLW}$ and $W$ giving the CEF energies $E_i$
according to eq. (\ref{eq:nuecross}).

\begin{widetext}
\begin{equation}
I_{E_{n}->E_{m}} = 4 \pi \left( \frac{\hbar \gamma e^{2}}{m c^{2}}
\right)^{2} \frac{exp(-E_{n}/k_{B}T)}{Z} \times \frac{2}{3} \sum
|<n|J|m>|^{2}. \label{eq:nuecross}
\end{equation}
\end{widetext}

An additional program was set up to fit the spectra using a wide
range of $x_{\rm LLW}$ and $W$ using a Lorentzian lineshape
broadening of the modes. To fit the spectral range close to origin
we also included an elastic peak.

Using the data collected on IN6 for six temperatures 5, 10, 15, 20,
30 and 50 K and using the estimate for $x_{\rm LLW}$ and $W$ as
obtained from the IN4 spectra  as starting values, we fitted the
spectra using the intensities and energies listed in Table
\ref{tm:prob}. Good fits over the whole temperature range up to 50K
were obtained for all temperatures with

\begin{equation}
x_{\rm LLW} = -0.44(1) \hspace*{.5cm}{\rm and}\hspace*{.5cm} W =
-0.222(2) {\rm{K}}. \label{eq:LLW}
\end{equation}

\begin{figure}
\includegraphics[bb = 3 4 511 249]{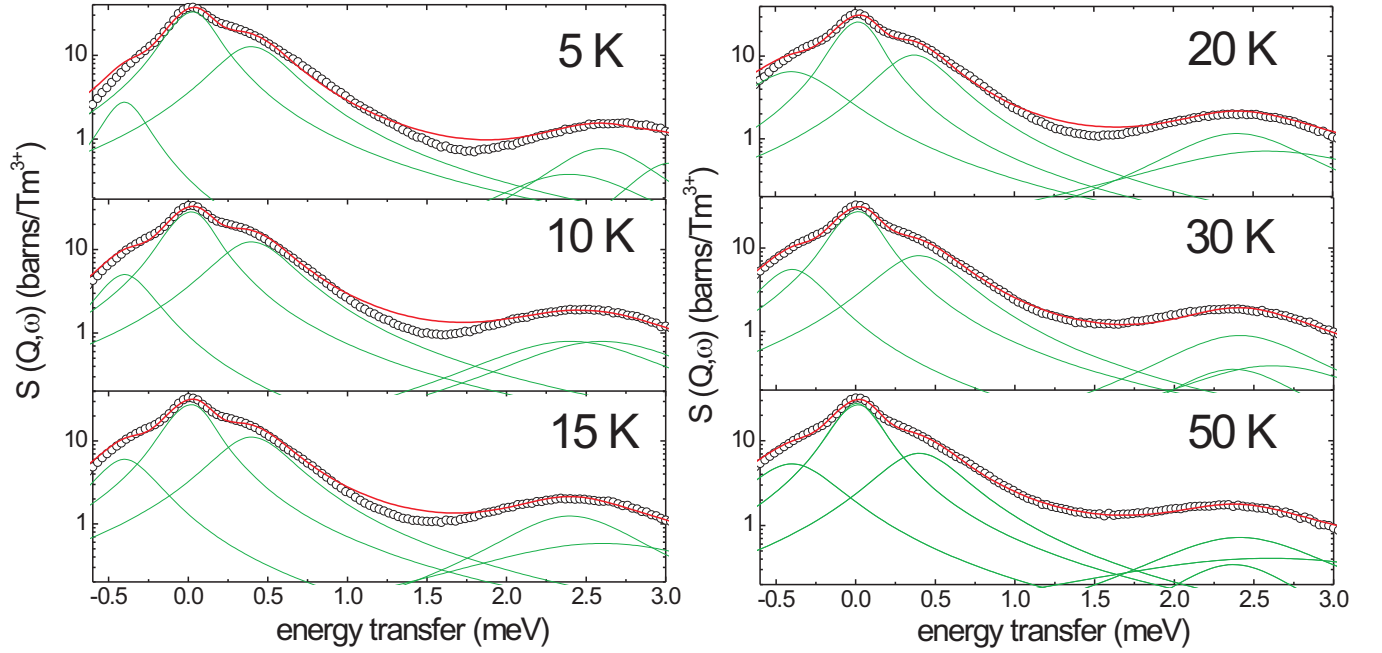}
\caption{(Color online) Semi-log plot of the IN6 spectra taken at 5,
10, 15, 20, 30 and 50 K. Black solid line represents that fit as
described in the text. The green solid lines are the individual
Lorentzian profiles.} \label{fig:IN6spectra}
\end{figure}

The fits for all temperatures taken on IN6 are shown in
fig~\ref{fig:IN6spectra}. Due to there being many transitions very
close to each other (see table~\ref{tm:prob}) the modes at 2.37 and
2.41 meV were merged into one mode. As well as the peaks at 2.59 and
2.61 meV and also the levels at $\pm$0.34 and at $\pm$0.40 meV. The
transition at 3.03 meV is increasingly intense at lower temperatures
(see table~\ref{tm:prob}) and was included into the fit at 5 and 10
K. Figure~\ref{fig:IN6spectra} gives the comparison of experiment
and calculated spectra.

\begin{table*} 
\caption{Transition probabilities according to (\ref{eq:nuecross})
for $x_{\rm LLW}$ = -0.44 and $W$ = -0.222 K of the crystal field
states of Tm$^{3+}$ in TmGa$_{3}$\label{tm:prob}}
\begin{tabular}{|p{2cm}|c|p{1cm}|p{1cm}|p{1cm}|p{1cm}|p{1cm}|p{1cm}|}

\hline
& & \multicolumn{5}{c}{$I_{E_{n}->E_{m}}$ (barn/ion)} &  \\
[5pt] $\Gamma_{n} \rightarrow \Gamma_{m}$ & $\Delta$ E (meV) & 5K &
10K & 15K & 20K & 30K & 50K    \\  [5pt] \hline $\Gamma_{2}$
$\rightarrow$ $\Gamma_{5}^{(2)}$ & 0.4 & 7.14 & 5.07 & 4.30 & 3.86 &
3.37 & 2.94 \\  [5pt]\hline $\Gamma_{1}$ $\rightarrow$
$\Gamma_{4}$ & 0.34 & 6.75 & 5.37 & 4.73 & 4.33 & 3.85 & 3.41 \\
 [5pt]\hline $\Gamma_{5}^{(2)}$ $\rightarrow$ $\Gamma_{2}$ & -0.4 & 2.85 &
3.20 & 3.17 & 3.07 & 2.89 & 2.68 \\  [5pt] \hline $\Gamma_{4}$
$\rightarrow$ $\Gamma_{1}$ & -0.34 & 3.08 & 3.63 & 3.64 & 3.56 &
3.38 & 3.15 \\  [5pt] \hline $\Gamma_{5}^{(2)}$ $\rightarrow$
$\Gamma_{3}$ & 2.41 & 1.35 & 1.52 & 1.50 & 1.46 & 1.37 & 1.27 \\
 [5pt] \hline $\Gamma_{5}^{(2)}$ $\rightarrow$ $\Gamma_{5}^{1}$ & 2.63 &
0.79 & 0.88 & 0.87 & 0.85  & 0.80 & 0.74 \\  [5pt] \hline
$\Gamma_{4}$ $\rightarrow$ $\Gamma_{3}$ & 2.37 & 0.44 & 0.52 & 0.52
& 0.51  & 0.48 & 0.45 \\  [5pt] \hline $\Gamma_{4}$ $\rightarrow$
$\Gamma_{5}^{(1)}$ & 2.59 & 0.91 & 1.07 & 1.08 & 1.05 & 1.00 & 0.93
\\  [5pt] \hline $\Gamma_{2}$ $\rightarrow$ $\Gamma_{5}^{(1)}$ & 3.03
& 1.34 & 0.95 & 0.8 & 0.72  & 0.63 & 0.55
\\  [5pt] \hline
\end{tabular}
\end{table*}

Further analysis of the transitions shows that at 30 and 50 K the
agreement with the fit profiles and the measurement is always better
than at lower temperatures. Especially at lower temperatures it
became increasingly difficult to fit the energy range between the
transition at 0.4 meV and 2.5 meV. One can see that at 20K the
agreement with the data and the fit is good for the transition
energies and the energy regions around them, but for the part of the
spectrum between 1 meV and 1.5 meV there is a gap between the fit
and the data, this persists from 20 to 5 K. The reason for this
discrepancy may be due to a quasielastic contribution which one
would expect to decrease  with increasing temperature.  A
quasielastic line was not included in the fits. Additionally, at
these low energies, there could also be some broadening due to
magnetic dispersion.

\begin{figure}
\includegraphics[bb = 16 13 284 225]{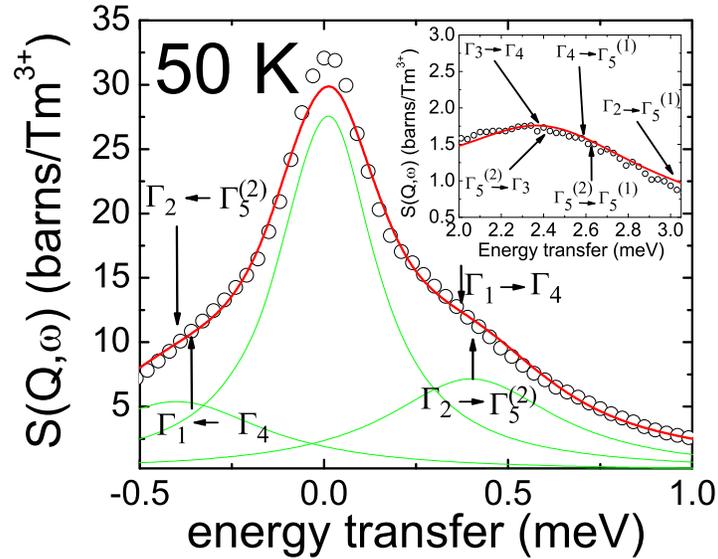}
\caption{(Color online) 50K spectra of TmGa$_{3}$. Crystal Field
transitions are shown for $x_{\rm LLW}$ = -0.44 and $W$ = -0.222 K
for the low excitations. Insert: 2.5 meV peak shown to consist of 4
modes. The red line is the fit obtained by fixing the intensities
and positions obtained from the program \textit{Mcphase} and using a
Lorentzian profile for each line.} \label{fig:IN6_50K}
\end{figure}

\begin{figure}
\includegraphics[bb = 16 18 295 226]{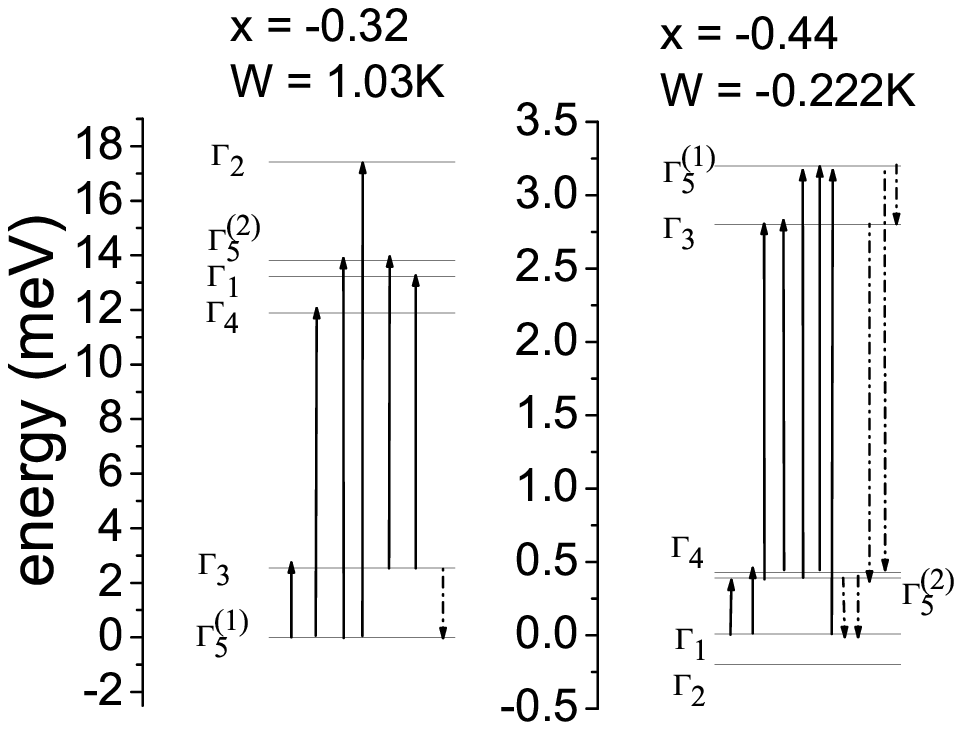}
\caption{left: Level scheme proposed by Morin \textit{et
al.}.\cite{Mo:tmnuetron} Right: Our level scheme, note that the
ground state ($\Gamma_{2}$) has been shifted down to -0.2 meV so to
be clearly seen from the $\Gamma_{1}$ which is only 0.009 meV away.
Solid arrows represent energy gain transfer, dash dot arrows energy
loss. Only the most intense transitions have been shown and for
$\Gamma_{n} \neq \Gamma_{m}$.} \label{fig:levelscheme}
\end{figure}

The 50 K spectrum  in comparison with the fit is displayed in more
detail in fig.~\ref{fig:IN6_50K}. The calculated energy positions of
the modes  according to Table~\ref{tm:prob} are also indicated. The
final CEF level scheme with the parameters given in
eq.(\ref{eq:LLW}) is shown in fig~\ref{fig:levelscheme} with a
comparison to the CEF level scheme proposed by Morin \textit{et
al.}. \cite{Mo:tmnuetron} The most intense transitions ($\Gamma_{n}
\neq \Gamma_{m}$) (cf. Table~\ref{tm:prob})  are shown for energy
gain and loss.

In order to reduce magnetic and quadrupolar and exchange interaction
between the Tm moments we studied the diluted Tm sample,
Tm$_{0.1}$Lu$_{0.9}$Ga$_{3}$ . Heat capacity measurements on these
sample showed no magnetic or quadrupolar ordering down to 1.8 K (see
below).

\begin{figure}
\includegraphics[bb = 15 18 291 221]{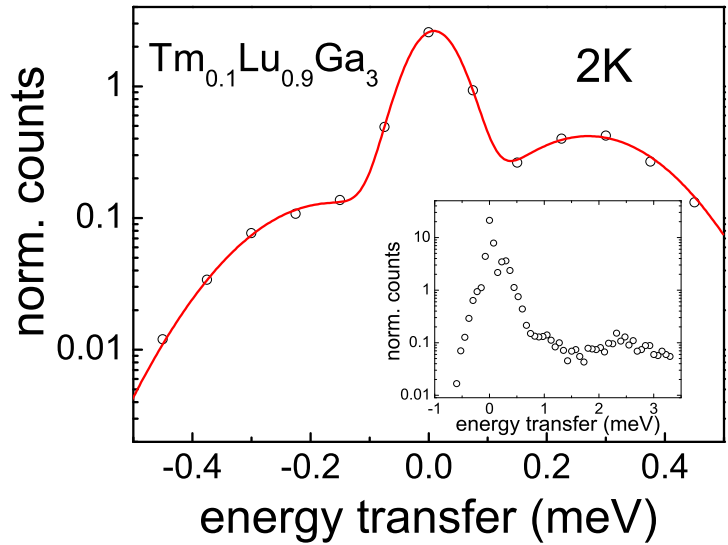}
\caption{(Color online) Semi-log plot of the IN12 data for
Tm$_{0.1}$Lu$_{0.9}$Ga$_{3}$ at 2 K. Open black circles are the data
points, solid line represents the fit as described in the text.
Insert: Full IN12 spectra from -0.6 meV to 3.5 meV.}
\label{fig:IN12}
\end{figure}

The spectrum at 2 K around the elastic peak measured on IN12 on the
diluted sample is shown in fig~\ref{fig:IN12}. The insert displays
the full spectrum at 2 K. The transitions at $E\sim$ $\pm$0.3 meV
can be clearly resolved being well separated from the elastic peak.
The mode at $\sim$2.6meV is also seen but it is as broad  as for the
concentrated  TmGa$_3$ sample. Additionally, there is a small
feature  at $\sim$1 meV which is not seen in the other data on IN4
on the Tm concentrated compound TmGa$_3$ and IN6. Data taken  at 10
K also showed similar features as those at 2 K, including the peak
at 1 meV. We tentatively ascribe  this extra feature to a splitting
of the $\Gamma_{5}^{(2)}$ third excited state. Such a splitting is
probably caused by some slight local deviations from cubic symmetry
due to the random substitution of the Lu sites by Tm atoms in the
diluted system Tm$_{0.1}$Lu$_{0.9}$Ga$_{3}$.  Such a splitting could
also be the reason why the mode at $\sim$2.6meV is as broad as in
the concentrated sample. By inspecting the relative intensities and
also the energy positions, we find slight deviations in $x_{\rm
LLW}$ and $W$ from the pure TmGa$_{3}$ to the diluted case, which
can be expected because of the slightly different lattice metrics of
LuGa$_3$ with respect to that of TmGa$_3$. Such a slight reduction
of the $W$ parameter for single ion CEF splitting of Tm in LuGa$_3$
and TmGa$_3$ is also confirmed by the analysis of the specific heat
data (see below).

The best fit of the CEF spectra of Tm$_{0.1}$Lu$_{0.9}$Ga$_{3}$ is
obtained with the parameter set

\begin{equation}
x_{\rm LLW} = -0.44 \hspace*{.5cm}{\rm and}\hspace*{.5cm} W = -0.18
{\rm K}. \label{eq:LLW}
\end{equation}

\subsection{Specific Heat and Magnetisation Measurements}

Our  INS results indicate a CEF scheme for TmGa$_3$ which is
significantly different from that proposed by Morin \textit{et al.}.
A characteristic feature of our CEF scheme is a separation of the
CEF levels into two groups about $\sim$2 meV apart. The low energy
group contains 4 CEF states ($\Gamma_2$ $\Gamma_1$,
$\Gamma_5$$^{(2)}$ and $\Gamma_4$) with an overall splitting of less
than $\sim$5K. The splitting of the CEF levels in the group around 2
meV ($\Gamma_3$ and $\Gamma_5^{(1)}$) is even smaller and amounts to
$\lesssim$ 2K. The small overall excitation energies of the CEF
level scheme of $\sim$25 K allow us to observe their contributions,
$C_{\rm m}$, to the total heat capacity at low temperatures and to
clearly distinguish these magnetic contribution from phonon
contributions which, at low temperatures, decrease rapidly according
to $C_{\rm ph} \propto T^3$. In order to determine the magnetic
contribution to the heat capacity, we measured the heat capacity of
samples with the composition Lu$_{1-x}$Tm$_x$Ga$_3$ ($x$=0.05 and
0.1) in the temperature range 1.8 - 40 K and of TmGa$_3$ in the
range 2 - 100 K and compared their heat capacities with the heat
capacity of LuGa$_3$, the latter representing the phonon
contribution. Samples with Tm diluted into the diamagnetic host
LuGa$_3$ allow to isolate - within limitations of cluster formation
with concomitant intercluster Tm - Tm interaction - the single ion
CEF behavior and to follow, by increasing the Tm content, the
effects of exchange and quadrupolar interaction.

\begin{figure}
\includegraphics[bb = 17 14 282 235]{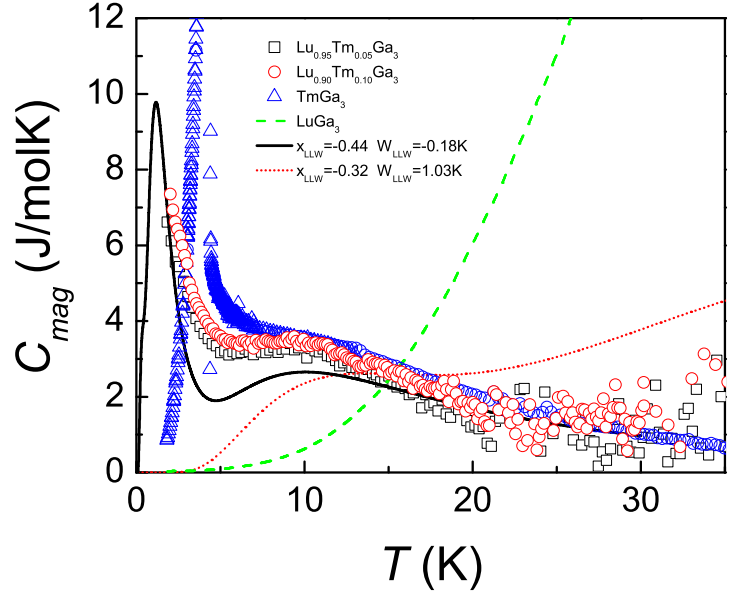}
\caption{(Color online) Magnetic contributions (per mole Tm atoms)
for Lu$_{0.95}$Tm$_{0.05}$Ga$_3$ (black $\Box$),
Lu$_{0.9}$Tm$_{0.1}$Ga$_3$ (red $\circ$) and TmGa$_3$ (blue
$\triangle$) obtained as described in detail in the text. The
(black) solid line represents the magnetic heat capacity calculated
for Tm$^{3+}$ in a cubic crystal electric field parametrized by the
Lea Leask Wolf parameters $x_{\rm {LLW}}$=-0.44 and $W$=-0.18K; the
dotted (red) line the magnetic heat capacity calculated for $x_{\rm{
LLW}}$=-0.32 and $W$=1.03K given by Morin \textit{et al.} (Ref.
\onlinecite{Mo:tmnuetron}) and the (green) dashed curve represents
the heat capacity of LuGa$_3$ taken as reference for the lattice
heat capacity.} \label{fig:CpSchott}
\end{figure}

Fig~\ref{fig:CpSchott} shows the resulting difference heat
capacities scaled by 1/0.05, 1/0.1 and 1 for
Lu$_{0.95}$Tm$_{0.05}$Ga$_3$, Lu$_{0.9}$Tm$_{0.1}$Ga$_3$ and for
TmGa$_3$, respectively. In order to compensate for small mismatches
of the high temperature data between Lu$_{0.95}$Tm$_{0.05}$Ga$_3$
and LuGa$_3$, the heat capacity of LuGa$_3$ was adjusted by the
factor 0.995 such that at high temperatures the resulting difference
magnetic heat capacities fall off $\propto 1/T^2$. Also shown in
Fig~\ref{fig:CpSchott} are the magnetic heat capacities calculated
from our CEF level scheme and for that given by Morin \textit{et
al.}.\cite{Mo:tmnuetron} In contrast to the expected scenario
according to the CEF scheme proposed by Morin \textit{et al.}
 our data reveal essential magnetic
contributions only up to about 30 K. They split into two
characteristic anomalies: A broad Schottky-type contribution, with
the maximum at $\sim$10K, followed by a sharp increase towards
lowest measuring temperatures. These features are very well
reproduced by the calculations using the CEF level scheme derived
from our INS measurements. By doubling the Tm concentration i.e.
going from Lu$_{0.95}$Tm$_{0.05}$Ga$_3$ to
Lu$_{0.9}$Tm$_{0.1}$Ga$_3$, the dip around $\sim$ 5 K between both
anomalies  is somewhat smeared out and the low temperature increase
becomes slightly more pronounced. TmGa$_3$ shows a very sharp
$\lambda$-like anomaly at the AFQ and AFM ordering temperature near
$\sim$4.27(1)K and a broad Schottky-like anomaly centered at
$\sim$10K, very similar to those observed for the diluted systems
but somewhat more smeared, with a long tail $\propto 1/T^2$ clearly
visible up to temperatures $\gtrsim$35K. It appears that the
magnetic entropy gained in the diluted systems below $\sim$5K for
TmGa$_3$ merged into the very sharp anomaly near the long-range
ordering point.

\begin{figure}
\includegraphics[bb = 12 15 285 222]{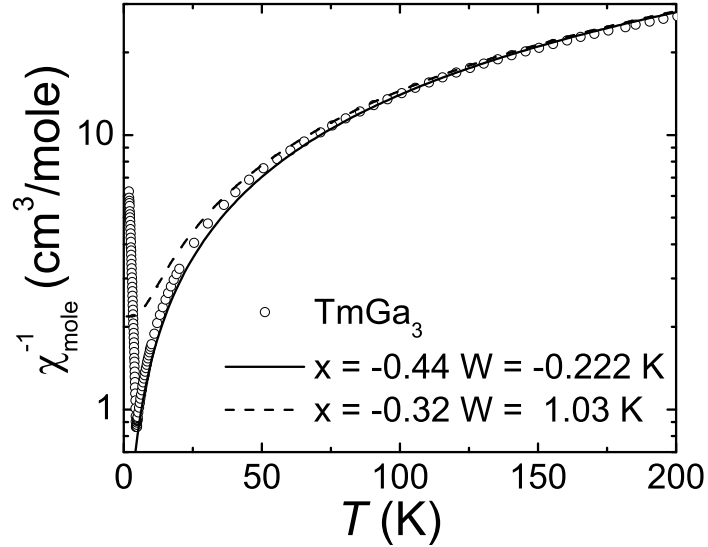}
\caption{Temperature dependence of the inverse Susceptibility of
TmGa$_{3}$. Open circles represent the data. Solid black line the
calculation using the new crystal field parameters. Dash line
calculation with the parameters of Morin \textit{et
al.}.\cite{Mo:tmmag}} \label{fig:suscept}
\end{figure}

Simulating  the temperature dependant susceptibility and the
magnetisation provides another way  of testing the CEF parameters.
Fig.~\ref{fig:suscept} shows a plot of the inverse susceptibility in
the range of 2 K to 200 K. The susceptibilties obtained from the CEF
parameters as obtained by our inelastic neutron spectroscopy agree
well with the experimental data above the transition temperatures.
Differences of the simulated susceptibilities using the CEF
parameters  $x_{\rm LLW}$ and $W$ given by Morin \textit{et
al.}\cite{Mo:tmmag} to the experimental data appear below $\sim$50
K. The good agreement of the fit lends further support to our CEF
scheme.

\begin{figure}
\includegraphics[bb = 7 0 270 396]{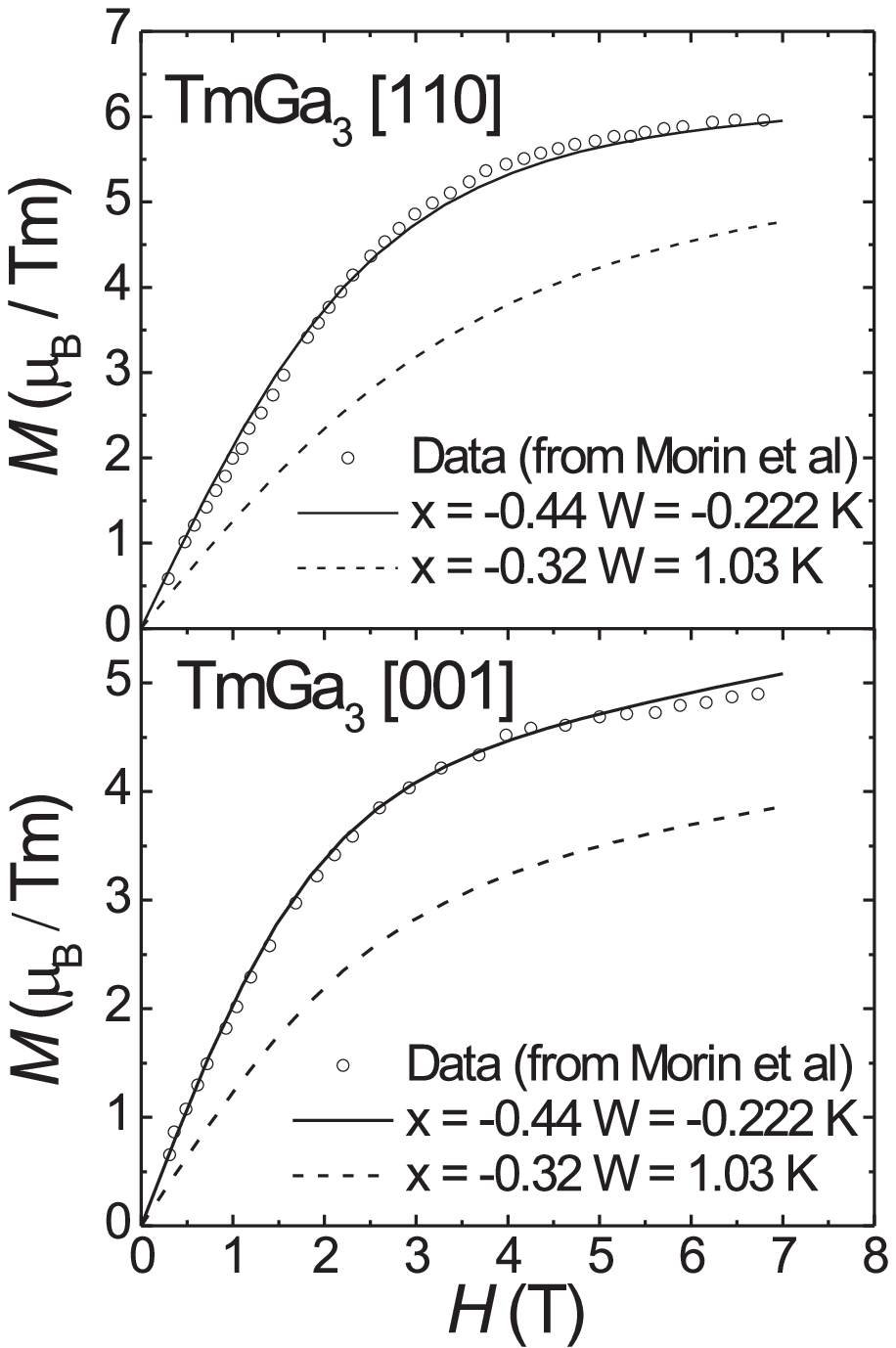}
\caption{Magnetisation at 10 K along Top: [110] Bottom: [001]. Data
taken from Morin \textit{et al.} (Ref. \onlinecite{Mo:tmmag}).}
\label{fig:m001}
\end{figure}

The magnetisation of TmGa$_3$ in the paramagnetic phase with
magnetic field applied along some main symmetry directions was also
calculated in order to check for consistency with our CEF model. The
results are shown in fig ~\ref{fig:m001} in comparison with data
collected at 10K by Morin \textit{et al.}  along [001] and [110]
directions.\cite{Mo:tmmag} We also display the magnetisations
calculated using the CEF scheme proposed earlier by Morin \textit{et
al.} which leads to significant discrepancies. In contrast, our CEF
scheme provides an almost perfect description of the magnetism of
TmGa$_3$ in the paramagnetic phase.

\subsection{Effect of a tetragonal distortion on the CEF level scheme}

\begin{figure}
\includegraphics[bb = 21 19 281 241]{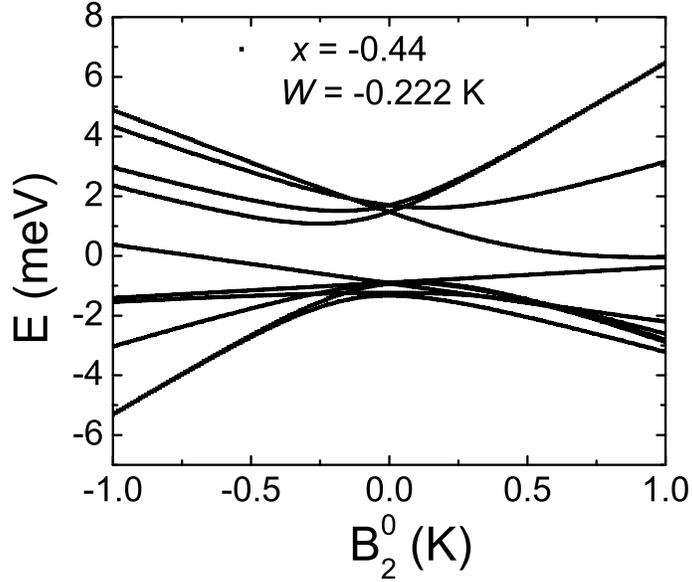}
\caption{Effect of a B$_{2}^{0}$ term in the Hamiltonian eq.
(\ref{eq:cubic}) on the CEF level scheme of TmGa$_{3}$.}
\label{fig:B20}
\end{figure}

The analysis of the heat capacity and the magnetisation of TmGa$_3$
and the samples Lu$_{1-x}$Tm$_x$Ga$_3$ provide very strong evidence
that the parameter set $x_{\rm LLW}$= -0.44 and $W$= -0.22 K
obtained from the INS measurements describe the CEF scheme of
Tm$^{3+}$ in TmGa$_3$ correctly. We now discuss possible effects of
the quadrupolar ordering on the CEF level scheme. Quadrupolar
ordering lowers the cubic symmetry and one has to consider
additional terms $B_{2}^{0}$$O_{2}^{0}$ and $B_{2}^{2}$$O_{2}^{0}$
in the CEF Hamiltonian (see eq. (\ref{eq:distortion})). If one
continuously turns on the $B_{2}^{0}$ term  a splitting of the
levels as shown in fig~\ref{fig:B20} results. The low energy states
become mixed already for small values of $B_{2}^{0}$ being of the
order of $B_{2}^{0}$ = -0.3 K as reported Gubbens \textit{et al.}
based on M\"ossbauer spectroscopy data.\cite{PG:moss}

\section{Discussion}

Our new set of crystal field parameters as obtained from inelastic
neutron scattering give thermodynamic properties which agree very
well with the experimental data. We have shown that the CEF scheme
proposed by Morin \textit{et al.} disagrees with thermodynamic
properties and should be discarded. As the crystal field energy
levels in TmGa$_{3}$ are very close to each other, and are all below
4 meV, determining such a level scheme is difficult. To ensure that
we have the correct parameter set, we have performed a number of
tests. First, using the results of Murasik for ErGa$_{3}$, by EPR
measurement on doped Er samples, we have confirmed that the level
scheme by Murasik \textit{et al.} is correct. We have then scaled
these values to give us an approximation of where in the parameter
space the correct values of $x_{\rm LLW}$ and $W$ for TmGa$_{3}$
could be. As neutron spectroscopy allows the best determination of
such parameters, we have fitted numerous spectra and determined
values that best suit all spectra. These values have been taken to
calculate thermodynamic properties, such as the Schottky anomalies
in the specific heat and the temperature dependant susceptibility as
well as the magnetisation in several lattice directions. The
agreement with the values obtained by inelastic neutron spectroscopy
and the thermodynamic properties assures us that we have obtained
the correct parameter set. Due to the small energy range of the
whole CEF level scheme AFQ ordering in TmGa$_{3}$ must involve an
essential  mixing of the CEF levels.  Our results can  be used for
further measurements in the ordered phase in order to understand the
complex and interesting competition between orbital and magnetic
ordering in TmGa$_3$.

\begin{acknowledgments}
We thank E. Br\"{u}cher for assistance with the SQUID measurements,
W. Schmidt for help with the IN12 measurements as well as M. Rotter
and K. U. Neumann for enlightening discussions.
\end{acknowledgments}

\bibliography{TmGa3}

\end{document}